# When Pricing Agents Meet Buying Agents: Personalized Pricing and Verifiable Trust

**Chupeng Xie**

## Abstract

Same fairness rule, different posterior, no trade. We study personalized pricing when seller and buyer principals delegate to agents that receive different value signals and execute machine-enforced mandates. Equal nominal surplus rules can be incompatible because each is applied to its agent's posterior. In one common environment, we solve nested pricing, inspection, mandate-selection, and repeated-relationship subgames. Signal attestation and execution attestation have different effects: evidence of a high seller signal can legitimate a high price, whereas evidence of a restrained rule can prevent unauthorized extraction. Verification can recover trade and soften both principals' policies only when it covers the disputed proposition. In repeated transactions, attributable offers above a buyer's reference cap depreciate relationship capital, producing a state-dependent Markov policy and a reference-respecting region. Verification is privately underprovided when the seller does not internalize avoided buyer inspection and relationship spillovers, but can be overprovided when it facilitates extraction. The model links AI measurement, algorithmic extraction, and verifiable restraint without treating trust as software emotion or verified execution as proof of true value.

## 1. Introduction

Two field experiments expose an intertemporal tension in data-driven pricing. In a machine-learning personalization study, Dubé and Misra (2023) estimate that personalized prices raise expected profit by an additional 19 percent relative to an optimized uniform price and reduce aggregate consumer surplus by 23 percent, although more than 60 percent of consumers receive a lower price. In a different setting, Anderson and Simester (2010) show that, among customers who had recently paid a higher price for an item later discounted, exposure to deeper discounts reduces subsequent orders by 14.8 percent over 28 months. The effect spills into other product categories and is largest among the firm's most valuable customers.

The treatments and markets differ, so the findings are not a logical contradiction. Together they identify a missing dynamic question. Better measurement can increase the current return to personalization, while an unfavorable price comparison can depreciate the relationship that supports future transactions. Existing models rarely place those forces inside the same pricing problem, and neither experiment studies what happens when both seller and buyer delegate valuation and price decisions to algorithms.

Personalized pricing has traditionally been modeled as a better-informed firm making an offer to a consumer. AI agents change both sides of the transaction. A seller can delegate valuation inference and price selection to a pricing agent. A consumer or procurement organization can delegate search, evaluation, negotiation, and acceptance to a buying agent. The transaction is then produced by two algorithms acting under policies selected by different human or organizational principals.

This institutional change is not a relabeling of the classical model. First, the two agents need not hold the same valuation estimate. The pricing agent may combine customer traces with population data, while the buying agent may observe private context supplied by its principal. Second, both agents are constrained delegates. A seller principal may limit discounts, require revenue targets, or prohibit the use of particular data. A buyer principal may impose a spending cap, a minimum estimated surplus, or a requirement that specified features of the pricing rule be verifiable. Third, each agent conditions on machine-readable identity, transaction history, and attestations. The equilibrium therefore depends on policy selection before it depends on price.

Recent economic work argues that AI agents can lower search and contracting costs while creating new questions about preference specification, identity, liability, and market design (Hadfield and Koh 2025; Shahidi et al. 2025). Emerging protocols similarly separate communication, authorization, payment, identity, and validation. Yet the personalized-pricing literature largely retains a passive or human buyer, while the agent-economy literature has not developed the pricing implications of two-sided algorithmic delegation. This paper connects the two.

We study a repeated relationship between a seller principal and a buyer principal. The seller first commits to an evidence regime. Conditional on that regime, the two principals simultaneously select executable mandates. In each period, latent value $v$ generates posterior means $m^S = E\left[v / D^S\right)$ and $m^B = E\left[v / D^B\right)$. The pricing agent selects an offer using $m^S$; the buying agent evaluates it using $m^B$, its policy, and evidence about the seller's signal and rule. A relationship distance records the consequences of prior attributable offers.

The first result is a policy-compatibility condition. Suppose the seller's declared rule promises to leave the buyer a share $\eta$ of value as estimated by the seller, so its rule price is $(1-\eta)m^S$. The buying agent requires a share $\sigma$ of value as estimated by the buyer and therefore accepts only prices no greater than $(1-\sigma)m^B$. Trade under the two mandates requires

$$(1-\eta)m^S \leq (1-\sigma)m^B.$$

Even if $\eta=\sigma$, trade fails whenever the seller's estimate exceeds the buyer's. Verification that the seller followed its rule does not eliminate disagreement about the relevant value. Agent markets therefore require both rule compatibility and an institution for resolving or bounding valuation disagreement.

The second result concerns information precision. A more precise seller signal improves targeting conditional on a passive buyer. With a buying agent, it also changes the distribution of posterior disagreement and can alter inspection and protective-policy choices under opacity. Its net return therefore need not be monotone: matching and price-targeting gains must be evaluated together with endogenous verification, rejection, mandate, and relationship responses. The formal model deliberately does not assign an unconditional sign to either signal precision.

The third result introduces relationship capital. An offer that exceeds the buying agent's reference cap raises relationship distance even when rejected, because it reveals a willingness to exploit policy or information asymmetry. The seller agent's dynamic program trades current extraction against lower future acceptance. We characterize a reference-respecting region using continuation-value losses rather than a steady-state shortcut. The region expands with expected relationship flow and credible identity persistence.

The fourth result concerns verifiable trust. We distinguish mandate verification, which proves the authority delegated to an agent; execution verification, which proves that a committed rule processed declared inputs; and outcome grounding, which links feedback to a transaction. No single proof establishes true valuation or full alignment with a principal's unexpressed intent. Verification reduces suspicion only over the propositions it covers. Because the instruments cover different stages, they can be complements rather than substitutes.

The model generates a policy-selection equilibrium. Opaque counterparties induce principals to program rigid agents: sellers restrict information release and buyers impose high surplus requirements. Stronger verifiability lets both sides delegate more discretion without creating equivalent exposure. This is the central institutional prediction: proof can mechanically restrict an execution while institutionally expanding discretion because separately designed optimizers must transact across organizational boundaries.

## 2. Related Literature

The behavior-based price discrimination literature studies customer recognition, commitment, and competitive targeting (Villas-Boas 1999; Fudenberg and Villas-Boas 2006; Li and Jain 2016). Work on superior knowledge shows how aggregate data can let a firm understand preferences better than customers do, creating suspicion, inspection, signaling, and a role for list prices (Xu and Dukes 2019, 2022; Li and Xu 2022). Aflaki and Zhang (2025) likewise show that price alone may not resolve uncertainty about whether personalization is being used and study an auxiliary information signal. We add a separately informed and strategically designed buyer agent whose own mandate and inspection choice feed back into price.

Research on price fairness and relationship dynamics shows that reference prices and antagonism affect demand (Kahneman et al. 1986; Anderson and Simester 2010). Zhang, Netzer, and Ansari (2014) estimate price-dependent latent trust states. We make that state an input into transactions between autonomous agents and connect it to portable identity and verifiable histories.

Strategic-delegation theory shows that principals may choose agents with objectives different from their own to obtain commitment or bargaining advantages (Fershtman and Judd 1987, 2006). AI makes agent policies more directly programmable but does not remove incomplete preference specification. Consistent with that distinction, Bichler, Gupta, and Oberlechner (2026) analyze auction markets in which automated bidding agents optimize delegated return-on-investment objectives rather than textbook quasilinear payoffs. Hadfield and Koh (2025) emphasize that a buying agent may imperfectly represent a human principal and that multi-agent interaction can magnify small behavioral differences. Our policy-selection stage applies this insight to price and surplus division.

Finally, the privacy and information literature studies consumer control, disclosure, and the welfare consequences of data use (Acquisti et al. 2016; Ali et al. 2023; Chen and Zhang 2026). We distinguish disclosure from evidence about delegated authority and executed policy. Atasu, Ciocan, and Désir (2024) study a seller principal that contracts with an agent to learn valuation and price; we add a separately informed buying agent whose mandate and evidence demand change the return to seller learning. Qiu et al. (2025) show that personalized ranking can raise prices through consumer search order; our transaction begins after a match reaches two delegated agents and asks whether their posterior-based policies are compatible. Den Boer and Keskin (2022) combine demand learning with reference effects; our relationship state is instead moved by an attributable offer relative to the buying agent’s cap and is carried by identity and transaction evidence. Eyster, Madarász, and Michaillat (2021) place fairness concerns in contemporaneous utility; we use a future switching and relationship channel rather than assuming a software agent experiences inequity aversion.

Two comparisons locate the model's incremental contribution. Fershtman and Judd (1987) study owners who publicly distort managers' incentives to change rival managers' actions in oligopoly. Here the strategic objects are bilateral, machine-enforced mandates chosen before trade; the delegates then receive different private signals, and verification changes which commitments the counterparty can condition on. Zhang, Netzer, and Ansari (2014) estimate vigilant and relaxed relationship states in B2B transactions and optimize prices against those states. We instead derive state movement from an attributable offer's excess over a buying agent's posterior cap, embed it in a two-agent policy game, and distinguish value-function existence from state-process stability. The contribution is not that the model adds one more feature to each literature. The interaction creates the paper's new mechanisms: nominally identical mandates can reject one another; signal evidence and rule evidence can move price and trust in opposite directions; and verification can restrict execution while relaxing the principals' chosen policies.

# 3. Model and Institutional Setting

## 3.1 Timing and delegated policies

The game uses one common primitive environment. At stage 0, the seller commits to an evidence regime $a$ and pays its fixed cost if adopted. At stage 1, conditional on that regime, the seller and buyer principals simultaneously select machine-enforced mandates. The seller chooses $M^S = \eta$, the promised buyer surplus share. The buyer chooses $M^B = \sigma$, its minimum estimated surplus share. The buyer's transaction-level evidence choice is determined endogenously by $e^*(p, b)$ in Section 4.2; an ex ante minimum-evidence requirement is reserved for the extension. The permitted seller data schema $D^S$ is fixed in the core model and determines the signal technology; endogenizing schema choice would require an additional information-acquisition cost and is reserved for future work. At stage 2, the agents receive private signals. At stage 3, the pricing agent offers a price and the buying agent accepts, rejects, or purchases evidence. At stage 4, the transaction record updates the relationship state. The evidence choice therefore anticipates the mandate equilibrium, while the two mandates are Nash choices within a fixed evidence regime.

The principals cannot costlessly override their policies after observing the agents' private signals. This commitment role distinguishes an executable agent from an advisory algorithm. Sections 4 and 5 solve the one-period pricing, inspection, and mandate subgames. Section 6 embeds the same price candidates and signal structure in the repeated relationship.

### 3.2 AI-measured value and mandate compatibility

Let latent match value be $v>0$. The seller and buyer agents observe information $D^S$ and $D^B$ and form positive posterior means $m^S=E\left[v/D^S\right)$ and $m^B=E\left[v/D^B\right)$. The seller's restrained rule and the buyer's hard cap are

$$p^R=\left(1-\eta\right)m^S, \tag{1}$$

$$c^B=\left(1-\sigma\right)m^B. \tag{2}$$

Here $\eta$ is the share the seller promises to leave according to the seller agent's estimate; $\sigma$ is the share the buyer principal requires according to the buying agent's estimate. They are economically comparable policy parameters but are applied to different information. Superior seller knowledge means lower expected prediction loss, not that $m^S=v$ realization by realization.

**Proposition 1. Bilateral compatibility**

Under hard mandates, a rule-compliant offer is feasible for the buying agent if and only if

$$\frac{m^S}{m^B}\leq\frac{1-\sigma}{1-\eta}. \tag{3}$$

If the agents share a common posterior, compatibility requires $\eta\geq\sigma$. If they use the same nominal surplus share, $\eta=\sigma$, compatibility requires $m^S\leq m^B$. Hence equal policy language does not ensure compatible execution.

The result separates a true policy conflict ($\eta<\sigma$) from a valuation disagreement and from a missing-evidence problem. Verification can resolve the third and sometimes the second; it cannot mechanically eliminate the first.

**Corollary 1. The disagreement tax**

Let $L = \log m^S - \log m^B$ have continuous distribution $F_L$. The probability that the restrained offer clears the buyer mandate is

$$F_L\left(\log \frac{1-\sigma}{1-\eta}\right). \tag{4}$$

A change in disagreement has no unconditional sign: it reduces trade when it shifts mass above the compatibility threshold and increases trade when it shifts mass toward the feasible side. Shared benchmarks and buyer-supplied proofs of a cap can therefore create value without changing either principal's preferred surplus division, but only in the relevant threshold region.

### 3.3 Evidence architecture and the relationship state

Evidence has four distinct objects: mandate coverage $a_M$ proves what the principal authorized; data coverage $a_D$ proves which declared input classes were included or excluded; execution coverage $a_E$ proves that committed code produced the offer; and outcome grounding $a_O$ links later feedback to the transaction. When a current pricing claim requires the first three stages, a serial benchmark is

$$a^{full} = a_M a_D a_E. \tag{5}$$

Signal evidence and execution evidence are not interchangeable. Let $z \in \{R, F\}$ denote the authorized rule or compliance state. For a buyer with signal $b$, the delegated action value after learning seller signal $s$ and rule state $z$ can be written

$$u(s, z; p, b) = (1-\sigma) E[v / s, b] - p - \chi_z(p), \chi_R(p) = 0, \chi_F(p) \geq 0, \tag{6}$$

where $\chi_F$ is the buyer principal's continuation or authority loss from an offer produced outside the protected rule. Signal attestation partitions uncertainty by $s$; execution attestation partitions it by $z$; joint attestation reveals both. A proof of high $s$ can legitimate a high price, whereas a proof of $z=R$ can rule out unauthorized extraction. Section 4 derives the separate values of these experiments and then solves a tractable signal-attestation benchmark with $z=R$ enforced.

Relationship distance is an economic state rather than a software emotion. Let $A_t \in \{0,1\}$ be realized identity attribution, with $Pr(A_t=1)=y$. Given the buyer's relevant posterior cap, the exact state transition is

$$r_{t+1}=min\{\acute{r},1+\rho(r_t-1)+\gamma A_t(p_t-c_t^B)_+\}. \tag{7}$$

The overage can be computed by a protocol against a committed cap or cap bin without revealing the cap itself. If no intermediary can produce that event, distance remains private to the buyer and the seller's state is a belief distribution, as discussed in the online appendix. Rejected offers can still affect the state when they are observable and attributable; no effect occurs when $A_t=0$.

For a fixed mandate pair, the pricing agent solves the common dynamic objective

$$V(r,I)=\max_{p\in P(M^S)} E[\pi(p,I,r)+\delta V(r',I')/I], \tag{8}$$

where $I$ is the seller's information state and $P(M^S)$ is the mandate-feasible action set. The seller's stage-0 evidence decision anticipates the policy equilibrium payoff $E_S^a$ and therefore satisfies

$$a^*\in arg\max_a\{E_S^a-K(a)\}. \tag{9}$$

This organization separates three objects that were previously easy to conflate: the agents' posterior values, their principals' executable policies, and the propositions that evidence can establish.

## 4. Bilateral Pricing and Inspection Equilibrium

The preceding sections state the economic forces in general form. This section closes the model and solves the pricing, acceptance, inspection, and delegation stages. The purpose is not to claim that commercial value is binary. The two-state economy is the smallest environment in which the agents can disagree for legitimate informational reasons, a price can signal the seller's information or conduct, and the seller can choose between intensive and extensive margins. Online Appendix E gives the continuous-signal extension.

### 4.1 Information and posteriors

Value is $v \in \{v_L, v_H\}$, $0 < v_L < v_H$, with prior $Pr(v = v_H) = \pi$. The pricing agent observes $s \in \{L, H\}$ and the buying agent observes $b \in \{L, H\}$. Conditional on value, signals are independent and have accuracies $\alpha_S, \alpha_B \in (1/2, 1)$. For each state index $i \in \{L, H\}$,

$$Pr(s = i / v_i) = \alpha_S, Pr(b = i / v_i) = \alpha_B. \tag{10}$$

Write $\ell_i = \alpha_i / (1 - \alpha_i)$ for the likelihood ratio of a high signal and $z_i = 2 1\{i = H\} - 1$. Bayes' rule gives the joint posterior odds

$$\frac{Pr(v_H / s, b)}{Pr(v_L / s, b)} = \frac{\pi}{1 - \pi} \ell_S^{z_s} \ell_B^{z_b}. \tag{11}$$

Let $\pi_{sb}$ denote the posterior probability implied by (11) and $m_{sb}=v_L+(v_H-v_L)\pi_{sb}$. The seller's posterior mean is $m_s=E[v/s]$. Conditional independence does not imply that the agents' signals are independent from one another: observing $s$ changes the seller's belief about $b$ through the common value. Specifically,

$$q_s \equiv Pr(b=H/s)=\alpha_B Pr(v_H/s)+(1-\alpha_B)Pr(v_L/s). \quad (12)$$

These expressions complete the signal structure omitted from reduced-form personalized-pricing models. They also make clear what an attestation can and cannot establish. Evidence that the seller received $s=H$ changes the buyer's posterior from $E[v/b]$ to $m_{Hb}$, but it does not make either signal equal to true value.

### 4.2 The buying agent's acceptance and evidence problem

The buyer principal programs a minimum surplus share $\sigma\in[0,1)$. If both signals are verifiably available, the buying agent accepts exactly when

$$p\le c_{sb}\equiv(1-\sigma)m_{sb}. \quad (13)$$

This rule is sequentially rational for an agent whose delegated payoff from buying is $(1-\sigma)m_{sb}-p$ and whose outside option is zero. More generally, suppose the pricing policy maps seller signal into price, $P:s\mapsto p_s$, and the buyer observes $(p,b)$ but not $s$. Its on-path belief is

$$v(s/p,b)=\frac{1\{P(s)=p\}Pr(s,b)}{\sum_{s'}1\{P(s')=p\}Pr(s',b)}. \quad (14)$$

When prices separate seller signals, price itself reveals $s$. When they pool, the buyer retains the posterior mixture in (14). At an off-path price we apply passive beliefs in the baseline—beliefs over $s$ remain $Pr(s/b)$—and use the D1 criterion as a robustness refinement. Online Appendix F states the additional type-level incentive constraints required for a separating, pooling, or semi-separating candidate; single crossing alone is not treated as sufficient.

The phrase *verification* can conceal two different objects. Signal evidence reveals $s$ and therefore changes the posterior over value. Execution evidence reveals whether the realized price complied with the committed rule $z$. Let $e \in \{0, S, E, SE\}$ denote no evidence, signal evidence, execution evidence, or both, and let $I_e$ be the corresponding information partition. The buying agent's gross value from evidence is

$$U^e(p,b) = E\left[max\{0, E\left[u(s,z;p,b)/I_e\right]\}/p,b\right) - k_e.$$

The general choice is $e^*(p,b) \in arg\,max_e U^e(p,b)$. Signal and execution evidence are substitutes when either one already determines the acceptance action, but they can be complements when only the joint pair identifies both value-relevant information and compliance.

The primitive evidence problem therefore separates signal and compliance uncertainty. For tractability, the remainder of Section 4 solves signal acquisition conditional on enforced compliance, while Section 5 endogenizes execution attestation through the principals' policy game.

For a closed-form benchmark, suppose execution evidence already enforces $z = R$ and the remaining question is whether to reveal $s$. Write the signal-inspection cost as $k_B$. Define

$$U^0(p,b) = max\{0, (1-\sigma)E[v/p,b) - p\}, \tag{15}$$

$$U^I(p,b)=\sum_s v(s/p,b)\max\{0,(1-\sigma)m_{sb}-p\}-k_B. \tag{16}$$

Inspection occurs if and only if $U^I(p,b)>U^0(p,b)$; ties are resolved against costly inspection. Acceptance after inspection follows (13), while acceptance without inspection follows the posterior cap $(1-\sigma)E[v/p,b]$. This is an equilibrium inspection rule rather than an exogenous "suspicion premium." Suspicion is the posterior probability that an observed offer is inconsistent with the restrained type or, equivalently, the posterior expected loss from acting on the pooled valuation.

**Lemma 1. Endogenous suspicion and the signal-inspection region**

For any $(p,b)$, define the gross value after inspection and the value without inspection as

$$A_B(p,b)=\sum_s v(s/p,b)\left[(1-\sigma)m_{sb}-p\right)_+,$$

$$B_B(p,b)=\left[(1-\sigma)E(v/p,b)-p\right)_+.$$

The value of inspection is therefore

$$\Delta_B(p,b)=A_B(p,b)-B_B(p,b). \tag{17}$$

The buying agent inspects exactly when $\Delta_B(p,b)>k_B$. The value is zero under a fully separating pricing policy, weakly positive under pooling by convexity, and strictly positive if learning $s$ changes the acceptance action with positive probability. Hence opacity creates inspection only in the disagreement interval

$$\min_s c_{sb}<p<\max_s c_{sb}. \tag{18}$$

The lemma supplies the Bayesian foundation for suspicion without conflating information with compliance. A high price need not increase signal inspection: if price fully reveals the high seller signal, additional signal evidence has no informational value. Execution evidence can still matter if the same price can be generated by compliant and noncompliant rules. Conversely, execution evidence has no decision value when $z$ is already known. Evidence demand concentrates where the unresolved proposition can change the action.

### 4.3 Static seller pricing equilibrium

First suppose the seller signal and compliance with the pricing rule are certified at zero marginal cost. Conditional on $s$, the pricing agent does not observe $b$ and faces two caps $c_{sL} < c_{sH}$. Any price below $c_{sL}$ leaves revenue on the table without increasing demand; any price strictly between the two caps serves only $b = H$ but is dominated by $c_{sH}$; any price above $c_{sH}$ produces no sale. The only relevant prices are therefore

$$p_s^A = c_{sL} \text{ and } p_s^T = c_{sH}, \tag{19}$$

where $A$ denotes broad coverage and $T$ targeted coverage. Expected revenues are $R_s^A = c_{sL}$ and $R_s^T = q_s c_{sH}$.

### Lemma 2. Seller pricing and market coverage

With verified signals and a hard buyer mandate, the unique revenue-maximizing price after seller signal $s$, except at equality, is

$$p_s^* = \begin{cases} c_{sL}, & q_s \le c_{sL} / c_{sH}, \\ c_{sH}, & q_s > c_{sL} / c_{sH}. \end{cases} \tag{20}$$

The first regime serves both buyer-signal types; the second serves only the high type. Increasing $\alpha_B$ widens the cap spread and can make targeting more attractive, while increasing $\alpha_S$ changes both $q_s$ and the two posterior caps. Lemma 2 therefore does not sign the effect of precision on the chosen coverage action. In the verified benchmark, a seller that can ignore a signal retains the usual value-of-information logic; any nonmonotone *net* return to precision must come from endogenous inspection, verification, mandate, or relationship responses elsewhere in the model.

Under opacity the same candidate-price principle applies, but demand is determined by (14)–(18). A pooling equilibrium can save the buyer's inspection cost while cross-subsidizing information states; a separating equilibrium transmits the seller signal through price but exposes the high-signal type to a stricter posterior response. Online Appendix F states the incentive-compatibility and refinement conditions for pooling, separating, and semi-separating candidates. The central point is that seller information affects not merely the conditional value distribution but the buyer's inference about the rule.

## 5. Mandate Choice and Verifiable Restraint

### 5.1 Principal policy selection

At stage 0 the seller selects evidence regime $a \in \{0,1\}$. Conditional on that observable regime, the seller and buyer principals simultaneously choose the buyer surplus share promised by the seller, $\eta$, and a minimum buyer surplus share $\sigma$. The seller promise imposes a price ceiling $\acute{p}_s(\eta) = (1-\eta) m_s$ when execution is certified. Let $K_S(a)$, $K_\eta(\eta)$, and $K_B(\sigma)$ be policy or implementation costs, with $K_S(0) = K_\eta(0) = K_B(0) = 0$. The seller's feasible candidate set is

$$P_s(\eta) = \{ min(c_{sL}, \acute{p}_s), min(c_{sH}, \acute{p}_s) \}. \tag{21}$$

Denote seller operating value by $\Pi(\eta, a; \sigma)$ and buyer surplus by $W(\sigma; \eta, a)$. For each observable evidence regime $a$, a selected mandate-subgame equilibrium $(\eta^a, \sigma^a)$ satisfies

$$\begin{aligned} \eta^a &\in \arg\max_{\eta} \{ \Pi(\eta, a; \sigma^a) - K_\eta(\eta) \}, \\ \sigma^a &\in \arg\max_{\sigma} \{ W(\sigma; \eta^a, a) - K_B(\sigma) \}. \end{aligned} \tag{22}$$

Let $E_S^a = \Pi(\eta^a, a; \sigma^a) - K_\eta(\eta^a)$ denote the seller's selected equilibrium payoff before the fixed evidence cost, under a stated selection rule if the mandate subgame has multiple equilibria. The stage-0 evidence choice is then

$$a^* \in \arg\max_{a \in \{0,1\}} \{ E_S^a - K_S(a) \}. \tag{23}$$

The buyer's policy cost includes false rejections caused by an overly cautious cap; the seller's restraint cost includes prices excluded by its promise. This makes both choices economically meaningful rather than free commitment devices.

### 5.2 Attestation choice

Using the selected mandate-subgame payoff defined above, seller attestation is adopted if and only if

$$E_S^1 - K_S(1) \geq E_S^0 - K_S(0). \tag{24}$$

If attestation eliminates duplicated inspection and prevents a positive mass of mistaken rejections, its adoption set expands with transaction frequency and the profit margin on recovered trades. It need not expand monotonically with $k_B$: when private inspection becomes prohibitively expensive, the buyer may reject rather than inspect, raising the seller's gain from attestation; when the buyer instead accepts the pooled offer, that gain can fall. The finite policy game below supplies explicit pure and mixed equilibrium conditions without relying on a generic continuous-action existence claim.

The last observation matters for interpretation. Verification is not automatically efficient, and its absence does not reveal technical infeasibility. It may reflect a distributional conflict: the seller does not internalize the buyer's avoided inspection cost, or the buyer does not internalize the seller's recovered margin.

### 5.3 Finite principal policy game

We now solve a finite version of the principals' first-stage game. Let each principal choose between a flexible and a protective mandate. The seller chooses $M_S \in \{F, R\}$. Under $F$ its agent may select either candidate price in (19); under $R$ it must use the broad-coverage price. The buyer chooses $M_B \in \{L, H\}$, corresponding to surplus requirements $0 \le \sigma_L < \sigma_H < 1$. The seller can attach execution attestation $a \in \{0, 1\}$ at per-relationship cost $K$. Without attestation, an $R$ claim is not self-proving; conditional on that claim, let $\mu$ be the public execution-risk prior that the flexible action is nevertheless used. With attestation, rule deviation is detected before purchase and the committed action is enforced. We first solve the conditional normal-form game for a fixed $\mu$ and fixed continuation payoff entries, then impose belief consistency when $\mu$ is generated by equilibrium mixing.

For each mandate pair, let $\Pi_{ij}^{a}$ and $U_{ij}^{a}$ be current-period seller operating profit and buyer surplus after solving the pricing and inspection subgame, where $i \in \{F, R\}$ and $j \in \{L, H\}$. Operating profit incorporates current acceptance and inspection responses but excludes the fixed attestation cost $K$ and the continuation relationship term defined below. These values are not primitives: they are computed from (10)–(20). Define the seller's flexibility gain

$$G_{j}^{a} = \Pi_{Fj}^{a} - \Pi_{Rj}^{a}, \tag{25}$$

and the buyer's protection gain

$$H_{i}^{a} = U_{iH}^{a} - U_{iL}^{a} - C_{H}, \tag{26}$$

where $C_H$ is the delegation-error and configuration cost of the stricter policy. The buyer chooses $H$ precisely when $H_{i}^{a} \geq 0$. The seller chooses $F$ when $G_{j}^{a}$ exceeds the relationship and commitment benefit of $R$. Let that benefit be $B_{j}^{a}$, equal only to discounted continuation or commitment value not already included in $\Pi_{ij}^{a}$, such as reduced future relationship distance. This accounting prevents current recovered trade from being counted twice.

**Proposition 2. Principal policy selection and attestation**

For a fixed attestation regime $a$, the policy game has the following pure equilibria:

1. $(R, L)$ if $B_{L}^{a} \geq G_{L}^{a}$ and $H_{R}^{a} \leq 0$;
2. $(R, H)$ if $B_{H}^{a} \geq G_{H}^{a}$ and $H_{R}^{a} \geq 0$;
3. $(F, L)$ if $B_{L}^{a} \leq G_{L}^{a}$ and $H_{F}^{a} \leq 0$;
4. $(F, H)$ if $B_{H}^{a} \leq G_{H}^{a}$ and $H_{F}^{a} \geq 0$.

If $H_R^a < 0 < H_F^a$ and $B_L^a < G_L^a$ while $B_H^a > G_H^a$, best responses cycle: flexibility induces protection and protection induces restraint. Conditional on a fixed payoff matrix, the unique equilibrium is then fully mixed. The seller chooses $R$ with probability

$$x^a = \frac{H_F^a}{H_F^a - H_R^a}, \tag{27}$$

and the buyer chooses $H$ with probability

$$y^a = \frac{G_L^a - B_L^a}{\left(G_L^a - B_L^a\right) + \left(B_H^a - G_H^a\right)}. \tag{28}$$

The proposition produces an interpretable institution-level equilibrium. Opaque flexible pricing makes the buyer protective; protection reduces the return to flexibility; credible restraint can then support a more permissive buying agent. The mixed region captures endogenous heterogeneity in policies when neither side can coordinate on a mutually flexible verified regime.

If the opacity belief is generated by the same seller randomization, Bayes consistency requires $\mu = 1 - x$. Then $H_i^0, G_j^0$, and $B_j^0$ can depend on $x$ through inspection and acceptance, and (27) becomes the fixed-point condition

$$x = \frac{H_F^0(x)}{H_F^0(x) - H_R^0(x)}. \tag{27A}$$

On an interval where the cycling inequalities hold, a continuous right-hand side mapping that interval into itself has a belief-consistent mixed equilibrium; it is unique if the mapping is a contraction. Without those restrictions, the opaque policy game can have multiple equilibria or a boundary solution. Thus (27) is a closed form for the conditional game, while (27A) is the equilibrium restriction when beliefs are endogenous.

Attestation is chosen before this policy subgame. With $K = K_S(1) - K_S(0)$, condition (24) is $E_S^1 - E_S^0 \geq K$. Because attestation changes the equilibrium probabilities $(x^a, y^a)$, its value includes a strategic-policy effect in addition to avoided inspection. Higher strict-mandate cost $C_H$ lowers both $H_i^a$ and shifts the buyer toward $L$. Holding the continuation function fixed, greater relationship sensitivity, attribution, or horizon raises $B_j^a$ whenever restraint lowers overage and the value function is decreasing; a lower fixed evidence cost expands the adoption set. These primitive shifts can alter equilibrium selection, so unrestricted global comparative statics are not claimed without the corresponding fixed-point restrictions.

### 5.4 Verification can soften both mandates

Suppose attestation raises $B_j$ by making restraint credible or lowers $H_R$ by eliminating the buyer's opportunism loss, without changing the true valuation distribution. Proposition 2 then permits a nonempty parameter region in which the opaque equilibrium is $(F, H)$ or mixed while the attested equilibrium is $(R, L)$. The next result verifies nonemptiness from the binary-signal primitives rather than assigning convenient values directly to the reduced-form payoff differences.

**Corollary 2. Primitive-supported policy softening**

Let $(v_L, v_H) = (1, 3)$, $\pi = .30$, $(\alpha_S, \alpha_B) = (.65, .90)$, $(\sigma_L, \sigma_H) = (.05, .25)$, and $C_H = .223$. For each buyer mandate, let $F$ choose the revenue-maximizing candidate in (20), let $R$ use the broad price, and let $X_j$ be the expected offer overage avoided by $R$. Set the continuation benefit to $B_j^a = \lambda \theta_a X_j$, where $\lambda = .15$ and the share of restraint made credible for continuation is $\theta_0 = .20$ under opacity and $\theta_1 = 1$ under attestation. Table 1 reports the resulting payoff differences and unique equilibria.

**Table 1. A primitive-supported shift from defensive mandates to verifiable restraint.**

| **Evidence** | $(G_L, G_H)$ | $(B_L, B_H)$ | $(H_F, H_R)$**; unique equilibrium** |
|---|---|---|---|
| Opaque | $(.0375, .0296)$ | $(.0109, .0086)$ | $(.0048, -.0031)$; $(F, H)$ |
| Attested | $(.0375, .0296)$ | $(.0545, .0430)$ | $(.0048, -.0031)$; $(R, L)$ |

Thus the attested institution uniquely shifts both principals from the defensive opaque outcome to credible restraint and the permissive buyer mandate. The selected seller payoff rises from .8543 to 1.0990 before the fixed evidence cost, so the stage-0 seller adopts whenever $K_S(1) - K_S(0) < .2447$. Every table entry is generated from the stated value distribution, signal structure, mandate costs, and continuation primitives by numerics.py; the online appendix records the calculation.

This result resolves an apparent paradox. A proof system is mechanically restrictive—it excludes noncompliant executions—yet institutionally it can increase flexibility because principals no longer need blunt protective rules. The empirical implication is that measuring only the number of rejected actions understates the benefit of verifiability; one must also measure how principals rewrite mandates after evidence becomes available.

### 5.5 Equilibrium discipline under opaque pricing

The inspection analysis permits pooling and separation. We now state conditions that discipline off-path beliefs. Consider two seller information types $s \in \{L, H\}$ choosing prices before the buyer acts. A pure separating equilibrium uses $p_L \neq p_H$ and makes (14) degenerate. It satisfies incentive compatibility if

$$\Pi_L(p_L; L) \geq \Pi_L(p_H; H), \Pi_H(p_H; H) \geq \Pi_H(p_L; L), \tag{29}$$

where demand incorporates the buyer's inference and inspection choice. A pooling equilibrium at $p^P$ requires neither type to gain from a deviation evaluated under an admissible off-path belief.

We apply the D1 criterion. Following an off-path upward price, beliefs place weight on the type with the larger set of buyer responses for which the deviation is profitable. Because a high seller signal raises the posterior distribution of buyer signal by (12), the high type generally has greater willingness to choose an upward deviation. D1 therefore assigns the deviation to $H$ when the single-crossing condition

$$\frac{\partial^2 \Pi_s(p)}{\partial p \, \partial s} > 0 \tag{30}$$

holds in the lattice sense. This makes upward price deviations less attractive as a way for the low type to mimic high value.

Under ordered posteriors, monotone buyer demand, and the stated single-crossing condition, a candidate least-cost separating equilibrium leaves the low type at its complete-information optimal price and assigns the high type the lowest price that both deters low-type imitation and respects the high buyer cap. The low price is broad when $q_L c_{LH} \leq c_{LL}$ and targeted otherwise, as in Lemma 2. Existence additionally requires the high type's own incentive constraint. A pooling candidate survives D1 only if every upward deviation attributed to the high type is unprofitable after the buyer's best response. A semi-separating candidate requires high-type indifference, low-type incentive compatibility, no other profitable deviation, and sequentially rational buyer behavior. If buyer mixing is needed to support it, buyer indifference is an additional condition and the buyer's mixing probability must enter seller profit. Under the baseline tie-break in Section 4.2, a buyer who is exactly indifferent does not inspect. Online Appendix F records these distinctions.

This discipline supplies a microfoundation for stochastic verification demand without claiming that every parameterization admits semi-separation. It also prevents an overstatement: greater price dispersion need not always cause more suspicion. Dispersion that cleanly separates legitimate information can reduce inspection, whereas overlap between pricing rules increases it.

**5.6 Numerical verification subgame**

Figure 3 solves a transparent policy comparison rather than imposing a behavioral suspicion coefficient. Under opacity, the seller principal commits to one pooled price across seller signals. Buyers observe their own signals, draw private inspection costs uniformly on $[0, \acute{k})$, and inspect exactly when the value in (17) exceeds the realized cost. The seller anticipates these decisions and reoptimizes the pooled price for every $(\acute{k}, \omega)$, where $\omega$ is the relationship-penalty weight. Under attestation, the seller instead reoptimizes signal-contingent prices from the candidate set in (19), and the buyer observes the certified seller signal before accepting. The calibration is $(\pi, \alpha_S, \alpha_B, \sigma) = (.75, .82, .80, .18)$; the plotted contour compares the gross seller value of attestation with fixed cost $K = .50$.

This restricted comparison is equilibrium-complete for the two commitment regimes just described, but it is not presented as the unrestricted signaling game. The code separately checks equation (17), seller price-grid convergence, and the Bellman residual. The resulting inspection surface has many interior rates because transaction-level inspection costs are heterogeneous, and the attestation panel reports the continuous value gain before drawing the adoption contour. Online Appendix F states the additional constraints required for pooling or semi-separation when the seller has no ex ante pooling commitment.

## 6. Dynamic Relationship Capital

We now embed the solved stage game in an infinite-horizon relationship. Each period draws a new value and new signals. The public state $r \in [1, \acute{r})$ is relationship distance; current monetary profit is divided by $r$, a parsimonious representation of lower conversion and greater search friction. Let $A_t \in \{0, 1\}$ indicate whether the offer is attributable to the persistent seller identity, with $Pr(A_t = 1) = y$. After price $p$ and buyer signal $b$,

$$r' = R_{sb}^{A}(r, p) = min\{\acute{r}, 1 + \rho(r - 1) + \gamma A(p - c_{sb})_{+}\}. \tag{31}$$

An excessive offer harms the relationship even if the low-signal buying agent rejects it, because an attributable offer reveals the seller’s willingness to exploit a valuation disagreement. The realized event $A_t$, rather than its expectation, updates the state. Holding the action-induced forcing fixed, the inherited component of relationship distance decays by the factor $\rho$ each period. A state-dependent policy can nevertheless create jumps in the induced transition map when its action changes.

Let $C_s$ be the two candidate prices in (19), adjusted for any seller mandate. The Bellman equation is

$$V(r) = \sum_{s} Pr(s) \max_{p \in C_s} Q_s(r, p), \tag{32}$$

where

$$Q_s(r, p) = \sum_{b} Pr(b/s) \left[ \frac{p 1\{c_{sb} \geq p\}}{r} + \delta\{(1 - y) V\left(R_{sb}^{0}(r, p)\right) + y V\left(R_{sb}^{1}(r, p)\right)\} \right).$$

The Bellman operator is a contraction on bounded continuous functions with modulus $\delta$. A unique value function and stationary Markov policy therefore exist. Because the action set is finite, the policy is fully characterized by the difference between targeted and broad coverage.

Existence of a unique value function concerns the optimization; it does not by itself imply that the state process induced by an optimal policy has a unique long-run distribution. The next lemma gives a sufficient condition for state stability. Because the optimal action can be set-valued at an indifference point, fix a measurable stationary optimal selector $p_s(r)$ before defining the induced process.

**Lemma 3. Conditional relationship-state stability**

For each realized triple $(s,b,A)$, let

$$T_{sbA}(r)=min\{\acute{r},1+\rho(r-1)+\gamma A(p_s(r)-c_{sb})_+\}$$

be the state transition induced by the fixed selector. Suppose primitives are independent across periods, let $\pi_{sbA}=Pr(s,b,A)$, each $T_{sbA}:[1,\acute{r}]\to[1,\acute{r}]$ is globally Lipschitz with constant $L_{sbA}<\infty$, and

$$\kappa:=\sum_{s,b,A}\pi_{sbA}L_{sbA}<1.$$

Then the induced Markov kernel $P$ has a unique invariant distribution $\mu^*$. For any two initial laws $v$ and $\eta$ on $[1,\acute{r}]$,

$$W_1(vP^t,\eta P^t)\le\kappa^t W_1(v,\eta).$$

In particular, $W_1(vP^t,\mu^*)\to 0$ geometrically, so long-run expectations of Lipschitz functions of relationship distance do not depend on the initial law.

The condition is automatic in a parameter region where the selected action is state-independent over the entire state space for every signal: the forcing term is then constant and the cap operator is nonexpansive, so $L_{sbA} \leq \rho$ for every realization. It is not automatic for a threshold policy. If the action-induced relationship increment jumps at a policy threshold, the global Lipschitz constant can be infinite even though the transition is affine with slope $\rho$ on either side. Single crossing therefore does not by itself establish state stability; threshold policies require a separate global-contraction, recurrent-class, or coalescence argument. The lemma makes no total-variation claim and does not identify $\rho$ as the process's exact convergence rate.

**Proposition 3. Markov pricing and the reference-respecting region**

Let $\Delta R_s(r) = [q_s c_{sH} - c_{sL}] / r$ be the current gain from targeting and let

$$\Delta C_s(r) = \delta \sum_b Pr(b/s) \sum_{A \in \{0,1\}} Pr(A) \{ V(R_{sb}^A(r, p_s^A)) - V(R_{sb}^A(r, p_s^T)) \} \tag{33}$$

be the continuation advantage of broad coverage. The stationary policy chooses the broad, reference-respecting price if and only if $\Delta C_s(r) \geq \Delta R_s(r)$. If the action difference $\Delta C_s(r) - \Delta R_s(r)$ is single crossing in $r$, the broad-price set is an interval (possibly empty or all of the state space). More generally, any parameter change that raises this difference pointwise expands the reference-respecting set. In particular, holding the continuation function fixed, greater $\delta$, $\gamma$, or attribution probability $y$ raises the relative value of broad coverage when $V$ is decreasing and the targeted transition is strictly higher with positive probability. Because these parameters also change the fixed point $V$, the unrestricted global comparative statics are reported numerically rather than asserted as a theorem.

This result separates two managerial levers. Increasing the reward horizon $\delta$ makes the pricing agent internalize future conversion; strengthening identity attribution $y$ makes current behavior follow the seller across protocol venues. Either can substitute partially for an explicit price constraint, but neither proves compliance with a promised rule.

**Numerical policy functions**

We solve (32) by value iteration on 401 equally spaced values of $r \in [1, 8]$, using linear interpolation for the next state. The baseline uses $(v_L, v_H) = (1, 3)$, $\pi = .5$, $\sigma = .2$, $\rho = .78$, and $y = .9$. Figure 1 reports posterior means and caps for $(\alpha_S, \alpha_B) = (.78, .72)$. They range from 1.20 after two low signals to 2.80 after two high signals; disagreement states generate intermediate means of 1.84 and 2.16. Thus equal nominal mandates do not create equal price caps.

Figure 2 uses $(\alpha_S, \alpha_B) = (.80, .90)$ and $\gamma = 1.5$ to display the economically relevant targeting region. At low $\delta$ the seller targets the high buyer signal. At high $\delta$ it serves both types. At an intermediate horizon the policy switches with relationship distance, demonstrating that the Markov state matters rather than merely relabeling a static discount factor. Figure 3 reports the equilibrium inspection rate and the gross seller value of attestation over $(\acute{k}, \omega)$; the white contour marks the fixed-cost adoption boundary. Figure 4 uses $\pi = .35$ and equal mandates $\eta = \sigma = .20$ to map compatibility and expected absolute posterior disagreement as the two agents' signal precisions vary. All figures are generated by the repository script numerics.py, which also writes the underlying numerical tables.

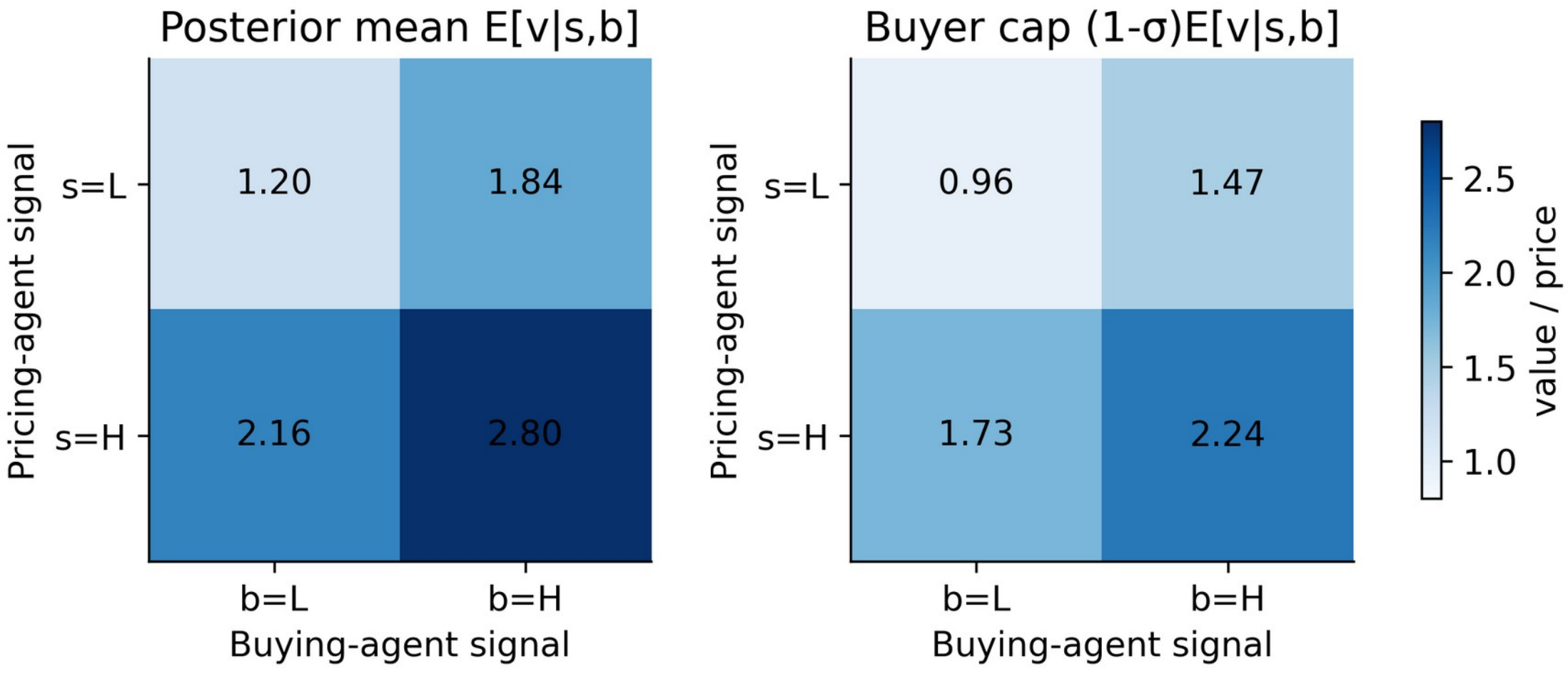

*Figure 1. Bilateral posterior means and buyer mandate caps.*

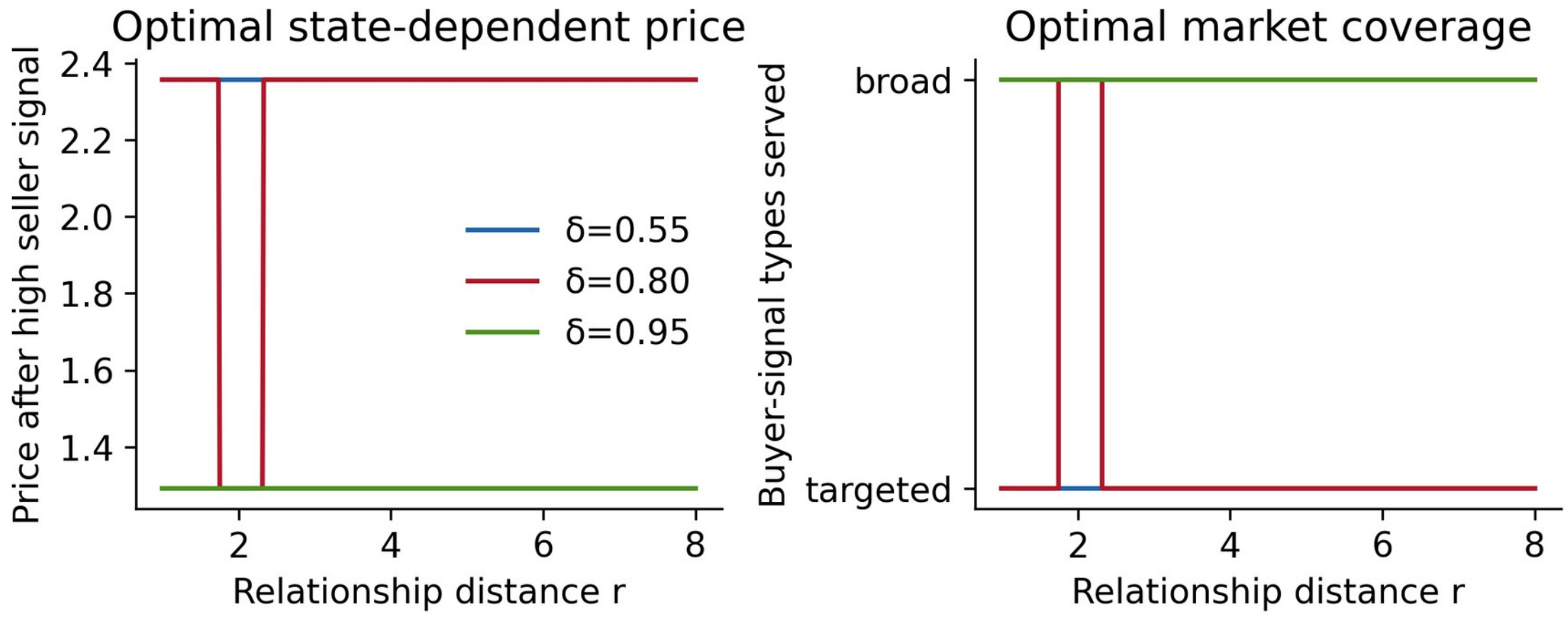

*Figure 2. Optimal dynamic price and market coverage.*

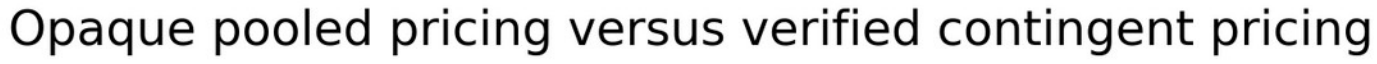

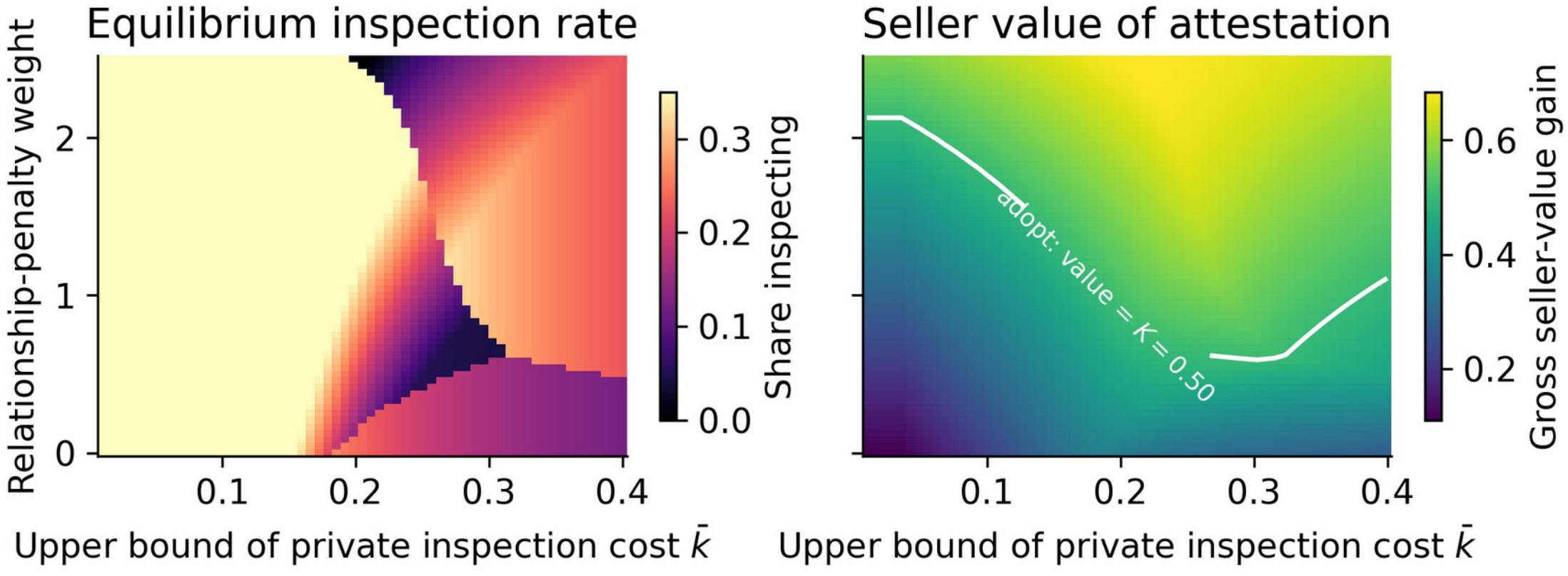

*Figure 3. Endogenous inspection and the gross seller value of attestation.*

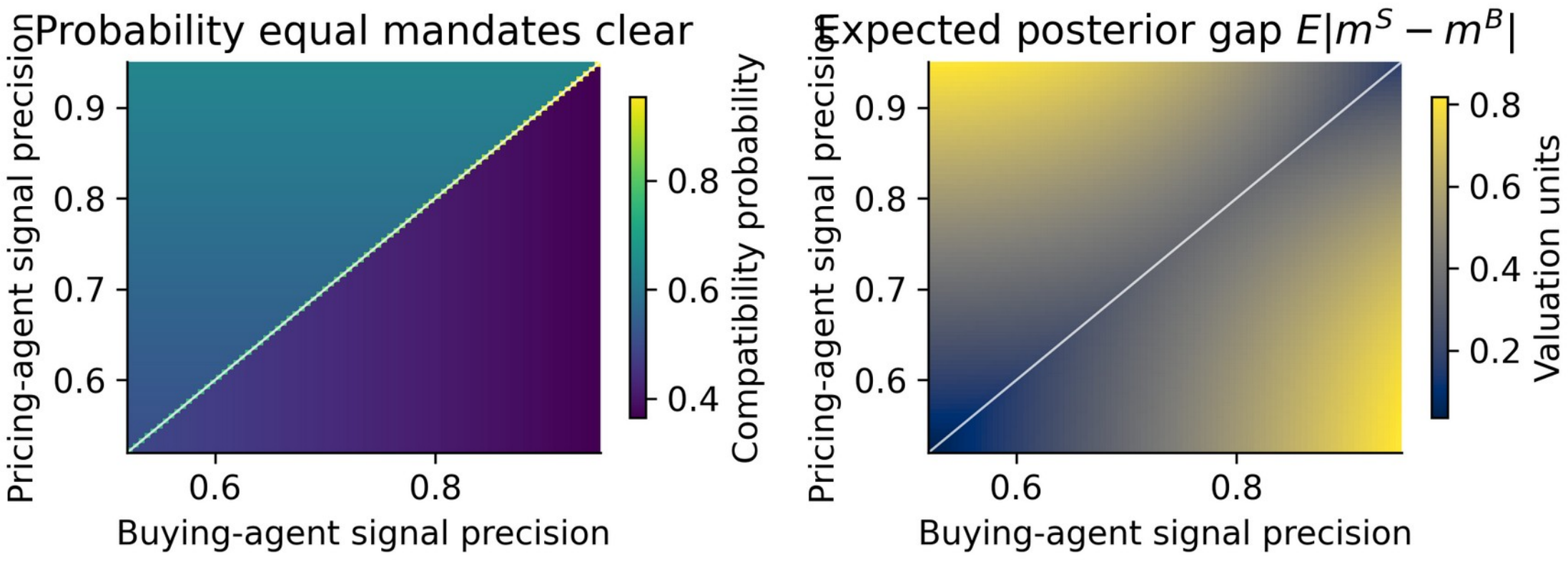

*Figure 4. Trade compatibility and posterior disagreement under equal nominal mandates.*

## 7. Verification Adoption and Welfare

Let expected gains from trade be $G_T = E[(v - c)1\{\text{trade}\}]$, expected inspection cost be $C_I = k_B Pr(\text{inspection})$, and verification cost be $C_V = K_S(a)$. Per-period total surplus is

$$TS = G_T - C_I - C_V - D(r), \tag{34}$$

where $c$ is marginal production cost and $D(r)$ is the real search and relationship friction associated with relationship distance. Price transfers cancel from total surplus but determine private incentives and continuation states. We compare four institutions: opaque personalization, ordinary rule disclosure, audited compliance, and full signal-and-execution attestation.

Ordinary disclosure changes beliefs only if false claims are costly. Audit makes claims probabilistically verifiable after the offer. Full attestation reveals the committed rule evaluation before acceptance but incurs a higher fixed cost. A list price is a special commitment that verifies a ceiling while forgoing some signal-contingent allocation.

**Proposition 4. Private versus socially optimal verification**

To study the intensive margin of verification, let $\alpha \in [0,1]$ denote continuous verification coverage, distinct from the binary evidence regime $a$ used in Sections 4–5. Let $G_S(\alpha)$ be the seller's profit gain from coverage $\alpha$, $G_B(\alpha)$ the buyer's avoided inspection and mistaken-decision loss, $G_R(\alpha)$ the reduction in future relationship friction borne by other sellers or protocol users, and $K_S(\alpha)$ the coverage cost. Assume these functions are differentiable, the private and social objectives are strictly concave, and the optima are interior; boundary optima satisfy the corresponding Kuhn–Tucker inequalities. Private seller coverage satisfies $G_S'(\alpha)=K_S'(\alpha)$.

Social coverage satisfies

$$G_S'(\alpha)+G_B'(\alpha)+G_R'(\alpha)=K_S'(\alpha). \tag{35}$$

At the private optimum, the planner's additional marginal return is $G_B'(\alpha)+G_R'(\alpha)$. If this sum is positive and the benefits are not transferred to the seller, private verification is inefficiently low; if it is negative, private verification is inefficiently high. Execution evidence for a restrained rule reduces mistaken rejection and private inspection, making $G_B'$ weakly positive in that comparison. Signal evidence that legitimates a higher accepted price without changing allocative efficiency can instead make $G_B'$ negative. Verification is therefore not a generic welfare improvement; its effect depends on the proposition proved and on the sum, not each component separately.

**Parameter regions**

The model yields four regions organized by the static targeting inequality and the dynamic-restraint inequality. When $q_s c_{sH} \le c_{sL}$, broad coverage is optimal even for a myopic seller. When the reverse inequality holds but (33) dominates, restraint is relational: it disappears if the seller is myopic or can reset identity. When targeting dominates both comparisons, the efficient response is not necessarily to prohibit personalized pricing; it may be to improve the buyer signal or enable the buyer to prove a cap without revealing its full valuation. Finally, if neither candidate price creates positive gains from trade for the low value state, no verification scheme should force trade.

Signal precision has two countervailing effects. Greater $\alpha_B$ improves buyer decisions but creates a wider cap spread, strengthening the seller's targeting incentive. Greater $\alpha_S$ improves allocation only if the seller signal predicts true value beyond what the buyer knows; it can otherwise increase price dispersion and inspection. Greater common data correlation reduces posterior disagreement but can create common-mode errors. These observations distinguish accuracy from institutional compatibility.

**Extensions and scope**

Online Appendix E extends the bilateral signal structure to continuous normal signals and shows that the seller prices against a distribution of delegated buyer caps rather than true value directly. Online Appendix G analyzes soft mandates, human escalation, two-seller verification adoption, private relationship states, portable reputation, and resettable identity. These extensions preserve the evidence-boundary and continuation-cost mechanisms while making clear which comparative statics require additional distributional or strategic-complementarity assumptions.

## 8. Empirical Predictions and Managerial Implications

This is an analytical paper, not an experimental-results paper. Online Appendix I gives a falsifiable identification design; it does not imply that uncollected evidence exists. The main predictions separate the mechanisms. Holding mandates fixed, trade falls with posterior disagreement rather than seller precision alone. Signal evidence reduces inspection only when it changes the buyer's decision estimate, whereas execution evidence reduces rejection only when uncertainty concerns compliance. Persistent identity and horizon interact through attributable continuation. Seller verification can fall below the social optimum when buyer inspection and relationship spillovers are external, yet exceed it when evidence mainly legitimates extraction.

Paired logs are essential: one-sided data confound posterior disagreement with buyer error or seller opportunism.

A controlled market can randomize signal precision, evidence type, and identity while principals choose payoff-bearing mandates. Field identification requires posterior bins, policy versions, offers, evidence, actions, outcomes, and persistent identities. Controlled offers identify $\gamma$ and $\rho$ through later conversion; randomized attribution identifies $y$; and disclosure-versus-attestation isolates information from compliance evidence. Committed intervals can preserve these comparisons without revealing exact valuations.

For sellers, a longer optimization horizon, a relationship-state transition, and a verifiable ceiling solve different problems. A pricing agent trained only on current margin omits the continuation cost in (33); giving that agent richer customer data can make extraction more accurate without making it more restrained. For buyers, $\sigma$ is a delegated insurance parameter: raising it protects against overpayment but increases false rejection when $m_{sb}$ is noisy. A layered mandate should therefore combine hard budget prohibitions, automatic acceptance in a high-confidence region, proposition-specific evidence requests, and escalation only when the unresolved value exceeds delay cost.

For marketplaces, the design object is an evidence vocabulary rather than one universal fairness rule. A transaction record should distinguish authorization, permitted data, signal provenance, rule execution, and settlement outcome. Google's Agent Payments Protocol illustrates the distinction by separating signed intent and cart mandates from payment execution. Such systems make authorization and transaction state machine-readable; they do not prove the buyer's true valuation or the normative fairness of a price. The model predicts value from standardizing what each evidence field establishes and from preserving attributable continuity across policy and model upgrades.

The scope is deliberately bounded. Buying agents here are executable delegated policies, not simulated human emotions. Relationship distance is a reduced-form state summarizing costly future search, rejection, and switching; Online Appendix G supplies a switching microfoundation. A proof can establish compliance with an encoded rule but cannot establish that $\sigma$ or $\eta$ is ethically correct. If the buyer remains passive, the model collapses to the one-sided benchmark. AI changes the economic object only when both sides form posteriors, commit to mandates, and condition execution on machine-readable evidence.

## 9. Conclusion

When pricing agents meet buying agents, price is the output of two inference systems and two delegated policies. The seller predicts what the buyer will pay; the buyer predicts what the purchase is worth and what the price reveals about the seller. Efficient trade requires compatible mandates as well as compatible valuations.

The analysis traces a three-part mechanism:

$$\text{AI measurement} \rightarrow \text{algorithmic extraction and response} \rightarrow \text{verifiable algorithmic restraint}.$$

Dynamic relationship capital can make restraint privately optimal. Verifiable evidence can distinguish restraint from cheap talk, but only inside a clearly defined boundary. At the policy-selection stage, proof supports rather than merely limits autonomy: principals can safely delegate more discretion when counterparties can verify how it was used.

The market-design challenge is consequently not to eliminate algorithmic discretion. It is to make separately designed agents legible and accountable to one another while preserving the informational gains that motivated delegation in the first place.

Proofs of Propositions 1–4, Corollaries 1–2, and Lemmas 1–3 appear in Online Appendix H. The one-sided dynamic benchmark, robustness analyses, opaque-pricing equilibrium conditions, auxiliary extensions, relationship-state microfoundation, and empirical design appear in Online Appendices A–I.

# Online Appendix to "When Pricing Agents Meet Buying Agents: Personalized Pricing and Verifiable Trust"

**Chupeng Xie**

## Appendix Overview

This appendix contains the one-sided dynamic benchmark, robustness analyses, continuous-signal and opaque-pricing extensions, auxiliary institutional results, proofs for the main paper, and a preregistration-ready identification design. It is part of the same submission and makes no claim of collected experimental evidence.

## Appendix A. One-Sided Dynamic Benchmark

### A.1 Information

A firm interacts repeatedly with a customer. The customer's persistent gross valuation is $v > 0$ and marginal cost is normalized to zero. Before pricing, the firm's AI system observes data $D^S$ and forms

$$\hat{v}^S = E\left[v / D^S\right).$$

The customer, or a purchasing agent acting for her, observes $D^B$ and forms $\hat{v}^B = E\left[v / D^B\right)$. The benchmark model sets $\hat{v}^S = \hat{v}^B = v$ to isolate the dynamic effect of relationship capital. Appendix A.6 restores disagreement and superior knowledge. A price is personalized because it is a function of $\hat{v}^S$ and the customer-specific relationship state.

## A.2 Demand and the relationship state

At date $t$, the firm offers $p_t \in [0, v)$. Let $r_t \geq \underline{r} > 0$ denote trust distance. In the linear benchmark, purchase probability is

$$q(p_t, r_t) = \frac{v - p_t}{r_t},$$

with parameters restricted so that $q \leq 1$. The firm's current expected profit is

$$\pi(p_t, r_t) = \frac{p_t(v - p_t)}{r_t}.$$

Appendix C gives a bounded logit specification, $q = (v - p)/(r + v - p)$, and shows that the qualitative results use only decreasing differences between price and relationship distance.

The customer regards a price as reference-respecting when it leaves her a share $\sigma \in (1/2, 1)$ of total surplus. The reference price is

$$\acute{p} = (1 - \sigma) v < v/2.$$

The restriction $\sigma > 1/2$ makes the reference price lower than the static monopoly price. It identifies the economically relevant case in which a fairness or delegation constraint can bind.

Trust distance evolves according to

$$r_{t+1} = R(r_t, p_t) = r_0 + \rho(r_t - r_0) + \gamma(p_t - \acute{p})_+, \tag{1}$$

where $r_0 > 0$ is baseline distance, $\rho \in [0,1)$ is persistence, and $\gamma > 0$ is relationship damage per unit of observable extraction. Equation (1) is offer-based: an observed offer can reveal an extractive rule even when rejected. This assumption is natural for digital and agent-mediated markets in which offers are logged. Appendix D considers purchase-weighted accumulation, $\gamma (p_t - \acute{p})_+ q_t$. That extension changes the quantitative boundary and is not treated as inessential.

### A.3 The firm's problem

For $n$ remaining periods, define

$$V_n(r) = \max_{p \in [0,v)} \{ \pi(p,r) + \delta V_{n-1}(R(r,p)) \}, V_0(r) = 0, \tag{2}$$

where $\delta \in [0,1)$. The infinite-horizon value $V$ is the fixed point of the corresponding Bellman operator. The firm observes the state and can choose a different price after trust changes.

### A.4 Dynamic personalized pricing

**Proposition 1. Existence, bounds, and state dependence**

For every finite horizon $n$, an optimal policy exists and $V_n(r)$ is continuous and weakly decreasing in $r$. The infinite-horizon Bellman operator is a contraction and has a unique bounded fixed point. If $\sigma > 1/2$, every optimal price satisfies

$$\acute{p} \leq p_n^*(r) \leq v/2. \tag{3}$$

In the final period, $p_1^*(r) = v/2$. Before the final period, the policy can depend on the current trust state through continuation value.

The bounds have a simple interpretation. A price below $\acute{p}$ produces the same next-period state as $\acute{p}$ and lower current profit. A price above $v/2$ produces lower current profit and weakly greater future distance than $v/2$. Both are dominated. The economically relevant choice is the extraction gap $x = p - \acute{p} \in [0, (\sigma - 1/2)v)$.

**Proposition 2. The reference-respecting region**

Let

$$r^b(r) = r_0 + \rho(r - r_0)$$

be next period's state when the firm respects the reference. With $n \geq 2$ periods remaining, $p_n^*(r) = \acute{p}$ if and only if, for every feasible extraction gap $x \in [0, (\sigma - 1/2)v)$,

$$\delta\left[V_{n-1}(r^b(r)) - V_{n-1}(r^b(r) + \gamma x)\right] \geq \frac{(2\sigma - 1)vx - x^2}{r}. \tag{4}$$

The left side is the discounted loss of relationship capital caused by extraction. The right side is its current profit gain. Equation (4) defines a state- and horizon-dependent reference-respecting region. It expands pointwise when a change in primitives raises the continuation loss from a given trust deterioration, including a longer remaining horizon or a larger weight on future profit.

For two remaining periods, $V_1(r) = v^2/(4r)$. Define

$$D(r, x) = \frac{1}{r^b(r)} - \frac{1}{r^b(r) + \gamma x}.$$

The condition becomes

$$\frac{\delta v^2}{4} D(r, x) \geq \frac{(2\sigma - 1)vx - x^2}{r} \text{ for all feasible } x. \tag{5}$$

A necessary local condition at the reference point is

$$\frac{(2\sigma-1)v}{r} \le \frac{\delta\gamma v^2}{4\left[r^b(r)\right]^2}. \tag{6}$$

At the baseline state, (6) is $\gamma \ge 4(2\sigma-1)r_0/(\delta v)$. Unlike a steady-state threshold, it correctly vanishes as a constraint when $\delta = 0$: a completely myopic firm chooses $v/2$.

**Corollary 1. Relationship scale**

Suppose the customer generates $m$ independent purchase opportunities per period, each affected by the same relationship state. Multiplying continuation flow by $m$ leaves the current extraction gain unchanged for a single focal offer and multiplies the future loss in (4). Thus high-frequency and high-cross-category customers enter the reference-respecting region at weaker antagonism. This is the model's precise connection to the finding that antagonism is especially costly among valuable customers.

**Discussion**

The result is a region, not a universal equality. Near the end of a relationship, after an anticipated identity reset, or when trust is already unlikely to recover, the firm may choose an interior price between $\acute{p}$ and $v/2$. With a long horizon and a valuable relationship, the reference can bind. The model therefore predicts both short-run extraction and long-run restraint without assigning different objectives to the firm.

## A.5 Robustness

### *A.5.1 Smooth reference responses*

Let trust accumulation be $\gamma h(p-\acute{p})$, where $h$ is nonnegative, increasing, and differentiable, rather than the kinked positive-part function. Exact bunching at $\acute{p}$ need not survive smoothing. The economic result does: the first-order condition includes the dynamic wedge

$$\delta \gamma h'(p - \acute{p}) V_r(R(r, p)) < 0,$$

which lowers the optimal price relative to a model that holds relationship state fixed. When $h'$ rises sharply around the reference, optimal prices concentrate in a neighborhood of $\acute{p}$. Thus the central prediction is a reference-respecting region or compression of extraction, not an artifactually exact equality under every specification.

*A.5.2 Heterogeneous references*

If $\sigma_i$ varies across customers with distribution $F_\sigma$, expected trust accumulation is smooth even when each individual has a kink. AI may predict not only valuation but also sensitivity to perceived extraction. Using the latter prediction to target customers creates a second-order commitment problem: the firm can exploit customers predicted to forgive quickly. A verifiable restriction on the use of behavioral-vulnerability features therefore differs from a conventional cap on price.

*A.5.3 Purchase-weighted antagonism*

Under $r' = r_0 + \rho(r - r_0) + \gamma(p - \acute{p})_+ q(p, r)$, trust accumulation is jointly determined with demand. For a stationary linear-demand benchmark, the steady state solves a quadratic rather than $r_\infty = r_0 + (\gamma / \lambda)(p - \acute{p})_+$. In particular, suppressing $q$ changes both the threshold and comparative statics. The dynamic programming characterization remains valid after replacing $\gamma x$ in (4) by $\gamma x q(\acute{p} + x, r)$, but monotonicity must be evaluated under the resulting transition.

### A.6 Information, suspicion, and superior prediction

The full-information benchmark isolates relationship capital. We now restore disagreement. The seller observes $D^S$ and the buyer observes $D^B$. A quoted price conveys information about both $\hat{v}^S$ and the seller's rule. Let $\mu$ be the buyer's posterior probability that the rule uses the seller's information opportunistically rather than respecting a declared restriction. Suspicion raises effective distance to

$$\tilde{r} = r\left[1 + s(\mu)\right], s(0) = 0, s'(\mu) > 0. \quad (7)$$

This reduced form can represent inspection, extra search, delayed acceptance, escrow, or a lower propensity to transact. It does not assume that customers directly observe the seller's estimate.

The distinction between $v$, $\hat{v}^S$, and $\hat{v}^B$ matters. A claim that the price leaves the buyer a share of true surplus is not verifiable when true $v$ is latent. A feasible commitment must instead refer to observable or attestable objects: a public list price; a declared benchmark; a cap relative to $\hat{v}^B$ supplied by the buyer agent; a committed model hash; an allowed input schema; or a rule applied uniformly to a declared class.

### A.7 Verifiable restraint

#### *A.7.1 Commitment technologies*

Index a commitment technology by verification coverage $a \in [0, 1)$, direct cost $k(a)$, and flexibility loss $\ell(a)$. Coverage is the fraction of economically relevant components of the declared rule that a customer can verify. The residual suspicion is $\mu(1 - a)$, giving

$$\tilde{r}(a) = r\{1 + s[\mu(1 - a)]\}. \quad (8)$$

List pricing typically has low cost and verifies a price ceiling but not feature use. Ordinary disclosure can describe feature use but has low coverage without enforcement. Audit provides probabilistic coverage and may occur after the transaction. Cryptographic or hardware-based attestation can verify execution of committed code on declared inputs, subject to data provenance, model-version, oracle, and implementation limitations.

For a fixed restrained price $p$ and relationship scale $A$, the firm's payoff from coverage is

$$U(a;A)=A\frac{p(v-p)}{r\{1+s[\mu(1-a)]\}}-k(a)-\ell(a). \tag{9}$$

**Proposition 3. Verification and relationship value**

Suppose $k+\ell$ is convex and the transaction-flow benefit in (9) is increasing in $a$. The optimal verification coverage $a^*(A)$ is weakly increasing in relationship scale $A$. If the solution is interior, it satisfies

$$A\frac{p(v-p)}{r}\frac{\mu s'[\mu(1-a)]}{\{1+s[\mu(1-a)]\}^2}=k'(a)+\ell'(a). \tag{10}$$

Verification is adopted first where transaction frequency, expected duration, or cross-category spillovers make suspicion expensive. This prediction differs from the claim that a particular cryptographic technology is universally cheap.

**Proposition 4. Separation through binding restraint**

Consider two privately observed firm types, $A_H > A_L$, that differ in the value of retaining the customer. Let $G(A)$ be the gross gain from adopting a specified binding commitment and suppose $G(A_H)>G(A_L)$. For a commitment cost $K$ satisfying

$$G(A_L)<K<G(A_H), \tag{11}$$

there is a separating perfect Bayesian equilibrium in which the high-relationship type commits and the low-relationship type does not. Customers assign probability one to the high type after observing valid commitment. If $K \leq G(A_L)$, both types may pool on commitment; if $K \geq G(A_H)$, neither adopts.

Binding restraint separates because it is more valuable to the firm that internalizes future transaction flow. Mere disclosure cannot produce the same equilibrium when both types can make the statement at negligible cost.

*A.7.2 Welfare*

At an unchanged price, reducing suspicion raises purchase probability and buyer surplus. Commitment is a Pareto improvement when the firm's recovered relationship value exceeds commitment and flexibility costs. If commitment also changes price, the welfare comparison must include price and participation effects; it is not universally positive.

**A.8 From human buyers to buying agents**

AI agents relocate rather than eliminate the frictions in the model. A human principal has underlying valuation $v$; a buying agent acts on $\hat{v}^B$; and a seller agent prices on $\hat{v}^S$. The buyer agent must infer both what the object is worth to its principal and how the offer was generated. A parameter such as $\sigma$ is then a delegated policy–for example, a minimum estimated surplus, a maximum markup relative to a benchmark, or a requirement for additional evidence before accepting an offer.

Trust distance becomes a machine-readable relationship state that can govern search, escrow, verification, and counterparty selection. Persistence is institutional rather than automatic. Authenticated, portable identity makes past conduct difficult to escape; cheap identity replacement, nonportable scores, model updates, and Sybil feedback weaken persistence. Thus $\rho$ is partly a protocol-design parameter.

The model yields four predictions for agent-mediated markets. First, persistent authenticated identities should reduce extraction when future rents are important. Second, reputation should influence transactions more strongly when feedback is grounded in verifiable transactions. Third, rule attestation should matter most when the seller has a large informational advantage. Fourth, verification should be adopted first in high-frequency and high-value relationships.

These predictions require protocol fields, not merely abundant logs. A useful empirical record contains the offer, public benchmark, committed rule version, verification coverage, acceptance decision, realized outcome, and subsequent counterparty choice. The resulting panel can separate immediate price sensitivity from a persistent relationship-state transition.

## Appendix B. Proofs for the One-Sided Benchmark

### Proof of Proposition 1

For finite $n$, the objective in (2) is continuous on the compact action set, so an optimizer exists. If $V_{n-1}$ is continuous, the maximum theorem gives continuity of $V_n$. Because current profit is decreasing in $r$ and $R(r, p)$ is increasing in $r$, induction implies that $V_n$ is weakly decreasing. The infinite-horizon operator discounts continuation value by $\delta < 1$, hence is a contraction in the sup norm.

For $p < \acute{p}$, the transition equals the transition at $\acute{p}$, while $p(v-p)$ is increasing on $[0, \acute{p})$ because $\acute{p} < v/2$; thus $p$ is dominated by $\acute{p}$. For $p > v/2$, current profit is lower than at $v/2$. Moreover $R(r, p) \geq R(r, v/2)$ and continuation value is decreasing in its state, so $p$ is dominated by $v/2$. In the final period continuation value is zero and the unique maximizer is $v/2$. QED.

**Proof of Proposition 2**

Write $p = \acute{p} + x$. Relative to $x = 0$, the current-profit difference is

$$\frac{(\acute{p}+x)(v-\acute{p}-x)-\acute{p}(v-\acute{p})}{r} = \frac{(2\sigma-1)vx-x^2}{r}.$$

The continuation-value difference in favor of respecting the reference is the left side of (4). Therefore $x = 0$ is globally optimal exactly when the continuation loss weakly exceeds the current gain for every feasible $x$. Equations (5) and (6) follow by substituting $V_1(r) = v^2/(4r)$ and differentiating at $x = 0$. QED.

**Proof of Proposition 3**

The payoff in (9) has increasing differences in $(a, A)$ because the transaction-flow term is increasing in coverage and multiplied by $A$, whereas costs do not depend on $A$. Monotone comparative statics imply that the greatest and least optimal selections are weakly increasing in $A$. Differentiating gives (10). QED.

**Proof of Proposition 4**

If the high type commits, its payoff gain net of cost is $G(A_H) - K > 0$. If it does not, it forgoes that gain. The low type would lose $K - G(A_L) > 0$ by mimicking. Given binding verification, a false commitment is infeasible. Beliefs placing probability one on the high type after valid commitment and zero after no commitment are Bayes-consistent on path. The pooling cases follow from the corresponding cost inequalities. QED.

## Appendix C. Bounded Demand

Let transaction utility be $\log[(v-p)/r] + \varepsilon_1$ and outside utility be $\varepsilon_0$, with independent type-I extreme-value shocks. Then

$$q(p,r) = \frac{v-p}{r+v-p}.$$

Current profit remains single-peaked. Prices below $\acute{p}$ have the same transition and a lower current payoff when the reference lies below the static optimum; prices above the static optimum lower current payoff and worsen the future state. Proposition 1 therefore extends with the static monopoly price replacing $v/2$. Proposition 2 extends by replacing the closed-form current gain with $\pi(\acute{p}+x,r) - \pi(\acute{p},r)$.

## Appendix D. Measurement-Error Extension

If the seller observes posterior density $f(v/D^S)$, expected current profit is

$$p\int_p^\infty q(p,r;v) f(v/D^S)\, dv.$$

The customer evaluates the offer using $f\left(v / D^{B}\right)$ and beliefs about the rule. A verifiable cap may be written against a public benchmark or a buyer-supplied posterior, but not against unknowable true $v$. Comparative statics in signal precision depend on market structure and are not signed in general: precision can improve matching, intensify extraction, or alter participation.

## Appendix E. Continuous Bilateral Signals

Let $v \sim N\left(\mu_0, \tau_0^{-1}\right)$, $s = v + \epsilon_S$, and $b = v + \epsilon_B$. With independent normal errors of precisions $\tau_S$ and $\tau_B$, conjugacy gives

$$m_{sb} = \frac{\tau_0 \mu_0 + \tau_S s + \tau_B b}{\tau_0 + \tau_S + \tau_B}.$$

$$Var\left(v / s, b\right) = \frac{1}{\tau_0 + \tau_S + \tau_B}.$$

Conditional on $s$, $m_{sB}$ is normal because it is affine in $B / S = s$. A verified mandate clears when $\left(1-\eta\right) m_s \leq \left(1-\sigma\right) m_{sB}$. Compatibility probability is therefore

$$1 - \Phi\left(\frac{\frac{1-\eta}{1-\sigma} m_s - E\left[m_{sB} / s\right]}{\sqrt{Var\left(m_{sB} / s\right)}}\right).$$

A mean-preserving increase in conditional posterior dispersion lowers compatibility when the threshold is above the conditional mean and raises it when below. Thus disagreement does not globally destroy trade; its sign depends on the mandate threshold’s location. With correlated errors, the same expressions use the inverse error covariance matrix. The main logic requires ordered posteriors, not conditional independence.

## Appendix F. Opaque-Pricing Equilibrium Conditions

This appendix makes explicit the equilibrium restrictions behind Section 5.5 of the main paper. Let the seller type be its signal $s \in \{L, H\}$, let $P_s$ be its possibly mixed price distribution, and let $d_s(p; v, k_B, \chi)$ be demand after the buyer updates to belief $v(s / p, b)$, uses receiver strategy $\chi(p, b)$ to inspect according to main-text equation (17), and then accepts or rejects. Seller type $s$ earns

$$\Pi_s(p; v, k_B, \chi) = p\, d_s(p; v, k_B, \chi). \tag{A12}$$

A separating candidate $(p_L, p_H)$ is a perfect Bayesian equilibrium only if beliefs are degenerate on path, buyer actions are sequentially rational, and both type-level incentive constraints hold. Define

$$D_L = \Pi_L(p_L; L, k_B, 0) - \Pi_L(p_H; H, k_B, 0),$$

$$D_H = \Pi_H(p_H; H, k_B, 0) - \Pi_H(p_L; L, k_B, 0).$$

The two constraints are equivalently

$$D_L \geq 0, D_H \geq 0. \tag{A13}$$

Under ordered posteriors and the lattice single-crossing condition in the main text, a least-cost separating candidate leaves the low type at its complete-information optimal price and chooses the smallest feasible $p_H$ that satisfies the first inequality in (A13). The low price is broad when $q_L c_{LH} \leq c_{LL}$ and targeted otherwise. The second inequality must still be checked; single crossing does not supply it automatically.

A pooling candidate $p^P$ uses the Bayes posterior $v(s / p^P, b) = Pr(s / b)$. It survives a specified off-path belief system only if, for each type,

$$\Pi_s\left(p^P; v^P, k_B, \chi^P\right) \geq \sup_{p \neq p^P} \Pi_s\left(p; v^p, k_B, \chi^p\right). \tag{A14}$$

Under D1, an off-path price is assigned to the type whose set of buyer responses making the deviation profitable strictly contains the other's. An upward deviation is therefore attributed to $H$ only when that profitable-response-set ordering holds; a cross-partial sign used as a convenient sufficient condition must be verified for the actual demand correspondence.

For semi-separation, suppose $L$ chooses $p^P$, while $H$ chooses $p^T$ with probability $\lambda$ and $p^P$ otherwise. At the pooling price, Bayes' rule gives

$$v_H\left(p^P / b; \lambda\right) = \frac{(1-\lambda) Pr(H, b)}{Pr(L, b) + (1-\lambda) Pr(H, b)}. \tag{A15}$$

Every semi-separating equilibrium requires high-type indifference,

$$\Pi_H\left(p^P; v^P(\lambda), k_B, \xi\right) = \Pi_H\left(p^T; H, k_B, 0\right), \tag{A16}$$

where $\xi = (\xi_L, \xi_H)$ records the buyer's inspection probabilities at an indifference history. The low type must not imitate $p^T$, neither type may have another profitable deviation under refined beliefs, and all acceptance and inspection decisions must be sequentially rational. If buyer mixing is required, then for each mixing buyer signal $b$, $\Delta_B\left(p^P, b; \lambda\right) = k_B$ and $\xi_b \in (0, 1)$; those probabilities enter the pooling-side payoff in (A16) through demand. Under the baseline tie-break in main-text Section 4.2, $\xi_b = 0$ at equality, so a semi-separating candidate must be supported without buyer mixing. Equations (A15)–(A16) show why buyer indifference by itself is never sufficient. Figure 3 deliberately solves the ex ante pooling-commitment subgame instead of reporting an unrestricted candidate that has not passed these checks.

# Appendix G. Auxiliary Extension Arguments

## G.1 Soft mandates

Replace the hard cap with acceptance probability $A(c-p)$, where $A$ is increasing and log-concave. The seller's current payoff becomes smooth, while a higher price still raises the next trust distance through expected excess over the reference cap. Discounting preserves the Bellman contraction. A monotone reference-respecting region follows when the ordered price action and next-state transition satisfy decreasing differences, the condition needed for the relevant Topkis comparison; it is not asserted from log concavity alone.

## G.2 Human escalation

Let review reveal an experiment that Blackwell-dominates the buyer's existing information. If the review and attestation experiments resolve the same seller-state uncertainty, replacing the former with costless certified information weakly raises the value of the accept/reject decision and eliminates the incremental value of paying for the redundant experiment. Escalation therefore weakly contracts. If review also reveals an orthogonal component of value, its residual value of information remains and can rise when attestation moves more transactions into the review-relevant region. This supplies the qualified escalation result summarized in main-text Section 7.

### G.3 Competitive adoption

Against a fixed rival action, adoption changes the seller's expected consideration-set demand by $\Delta D$ and yields incremental margin $m\Delta D-K$. Therefore no adoption is a best response to an opaque rival when $m\Delta D(0)\leq K$, and adoption is a best response to an attesting rival when $m\Delta D(1)\geq K$. If $\Delta D(1)\geq \Delta D(0)$, adoption decisions are strategic complements and both symmetric outcomes can coexist for intermediate costs. If the inequality reverses, intermediate costs can instead support asymmetric adoption. These are best-response conditions, not a claim that competition universally increases verification.

### G.4 Alternative relationship technologies and microfoundation

If trust distance is private to the buyer, the seller filters it from purchase and inspection behavior and the pricing problem becomes a partially observed Markov decision process. If reputation is portable across buyers, an attributable excess offer creates a state cost proportional to future market reach, expanding the private restraint term while also magnifying the cost of erroneous negative evidence. If identity can reset at cost $k_I$, the enforceable continuation penalty is capped by the lower of the lost reputation rent and reset cost. These cases yield distinct empirical signatures: within-buyer history dependence under private trust, cross-buyer spillovers under portable reputation, and discontinuities around identity migration under resettable identity.

A simple switching model microfounds the reduced-form distance. After observing an attributable offer history $h_t$, a buyer principal chooses whether to keep the counterparty or pay search cost $c_t$ to sample an alternative. Let the utility difference favoring retention be $\zeta - r(h_t)$, where $\zeta$ has a logistic distribution. Retention probability is then $\left[1+\exp\left(r(h_t)\right)\right]^{-1}$ up to location and scale. Around any interior baseline, dividing current transaction flow by the normalized index $r_t \geq 1$ is the first-order local representation of this decreasing retention probability. If an attributable over-cap offer shifts the index by $\gamma\left(p - c_{sb}\right)_+$ and ordinary experience depreciates the index at rate $\rho$, equation (31) is the Markov projection of the switching process. Thus $r_t$ need not be an emotion inside software: it can summarize principal-programmed search, avoidance, and reputation responses.

# Appendix H. Proofs for the Main Paper

## Proof of Main-Paper Proposition 1 and Corollary 1

Substituting main-text equations (1) and (2), the buying agent accepts the restrained price if $(1-\eta)m^S \leq (1-\sigma)m^B$. Dividing by positive $(1-\eta)m^B$ yields the compatibility ratio. The common-posterior and equal-share cases follow directly. Taking logs gives $L \leq \log\left[(1-\sigma)/(1-\eta)\right]$; applying the distribution function of $L$ gives the trade probability. QED.

**Proof of Main-Paper Lemma 1**

Without inspection, sequential rationality gives main-text equation (15). Inspection permits the action to depend on seller state and gives equation (16). Subtraction yields equation (17). Because $x \mapsto (x-p)_+$ is convex, Jensen's inequality makes gross information value nonnegative. It is strictly positive exactly when posterior realizations lie on opposite sides of the acceptance threshold. A separating price makes $v$ degenerate; under pooling, opposite decisions can occur only when the price lies between the two state-contingent caps. QED.

**Proof of Main-Paper Lemma 2**

Fix $s$. Demand is one weakly below $c_{sL}$, $q_s$ above $c_{sL}$ and weakly below $c_{sH}$, and zero above $c_{sH}$. Revenue increases with price within each interval. The maximizer is therefore either $c_{sL}$ or $c_{sH}$. Comparing $c_{sL}$ with $q_s c_{sH}$ establishes the result, uniquely except at equality. QED.

**Proof of Main-Paper Proposition 2**

For fixed attestation regime $a$ and a fixed subgame payoff matrix, compare the seller's payoff from $F$ and $R$ against each buyer action. The seller selects $R$ exactly when $B_j^a - G_j^a \geq 0$. The buyer selects $H$ exactly when $H_i^a \geq 0$. Combining these best-response inequalities yields the four pure cases. In the cycling region, let $x$ be the probability of $R$. Buyer indifference requires $x H_R^a + (1-x) H_F^a = 0$, which gives main-text equation (27). Let $y$ be the probability of $H$. Seller indifference gives equation (28). Both probabilities lie strictly inside the unit interval under the stated cycling inequalities.

If opacity beliefs are generated by the same mixing, Bayes consistency sets $\mu = 1 - x$, so the payoff differences become functions of $x$ and buyer indifference gives main-text equation (27A). On a compact cycling interval, continuity and the self-map condition give a fixed point by the intermediate-value/Brouwer argument; a contraction gives uniqueness. The conditional closed form alone is not asserted to solve that endogenous-belief case. Finally, comparing the seller's selected gross equilibrium payoffs under $a = 1$ and $a = 0$ gives the necessary and sufficient adoption cutoff $E_S^1 - E_S^0 \geq K$. QED.

**Proof of Main-Paper Corollary 2**

Bayes' rule under the stated value and signal primitives gives the buyer-cap matrices

$$C^L = \begin{pmatrix} .997500 & 2.232500 \\ 1.104375 & 2.617250 \end{pmatrix}, C^H = \begin{pmatrix} .787500 & 1.762500 \\ .871875 & 2.066250 \end{pmatrix}.$$

Rows index the seller signal and columns the buyer signal. Applying equation (20), the flexible policy uses the broad price after $s = L$ and the targeted price after $s = H$ under both buyer mandates. Direct integration over $(v, s, b)$ gives seller profits

$$(\Pi_{FL}, \Pi_{RL}, \Pi_{FH}, \Pi_{RH}) = (1.082050, 1.044525, .854250, .824625)$$

and buyer surpluses before the strict-policy cost

$$(U_{FL}, U_{RL}, U_{FH}, U_{RH}) = (.238950, .555475, .466750, .775375).$$

Therefore $(G_L, G_H) = (.037525, .029625)$ and, after subtracting $C_H = .223$, $(H_F, H_R) = (.004800, -.003100)$. The flexible policy creates expected attributable overages $(X_L, X_H) = (.363090, .286650)$ relative to broad pricing. Hence $B_j^a = .15\,\theta_a X_j$ gives $(B_L^0, B_H^0) = (.0108927, .0085995)$ and $(B_L^1, B_H^1) = (.0544635, .0429975)$.

Under opacity, $B_H^0 < G_H$ and $H_F > 0$, so Proposition 2 makes $(F, H)$ the unique pure equilibrium; the inequalities for the other three profiles fail strictly. Under attestation, $B_L^1 > G_L$ and $H_R < 0$, so $(R, L)$ is uniquely selected. The corresponding seller payoffs before the fixed evidence cost are .854250 and $1.044525 + .0544635 = 1.0989885$. Their difference is .2447385, which proves the stage-0 adoption cutoff. The repository CSV retains full-precision values and the numerical script asserts all strict equilibrium inequalities. QED.

**Proof of Main-Paper Lemma 3**

Let $(R, R')$ have any coupling of initial laws $\nu$ and $\eta$, and draw the same next-period realization $(S, B, A)$ for both states, independently of $(R, R')$. The synchronous coupling and global Lipschitz bounds give

$$E\left[\left|T_{SBA}(R) - T_{SBA}(R')\right| / R, R'\right] \leq \sum_{s,b,A} \pi_{sbA} L_{sbA} \left|R - R'\right| = \kappa \left|R - R'\right|. \tag{A17}$$

Taking expectations and the infimum over couplings yields $W_1(\nu P, \eta P) \leq \kappa W_1(\nu, \eta)$. Probability measures on the compact interval $[1, \acute{r}]$ form a complete space under $W_1$, so the kernel is a contraction. Banach's fixed-point theorem supplies a unique invariant distribution and iteration gives the stated geometric bound. QED.

**Proof of Main-Paper Proposition 3**

For bounded $V, W$, the maximum inequality applied to the Bellman operator gives $\| TV - TW \|_\infty \leq \delta \| V - W \|_\infty$, so a unique bounded fixed point and a stationary optimal selector exist. Comparing the targeted and broad actions and integrating over $(b, A)$ gives broad coverage exactly when the continuation advantage in main-text equation (33) weakly exceeds the current targeting gain. If their difference is single crossing in $r$, its weakly positive set is an interval. Any parameter change that raises the action difference pointwise expands that set. Holding $V$ fixed, increasing $\delta$ raises the continuation term; increasing $\gamma$ or $y$ raises the targeted next state with positive probability while leaving the broad next state unchanged because the broad price creates no excess over either buyer cap. Since $V$ is decreasing, the fixed-continuation comparison moves toward broad coverage. The proposition deliberately leaves unrestricted fixed-point comparative statics to the numerical analysis. QED.

**Proof of Main-Paper Proposition 4**

Under the differentiability, concavity, and interiority assumptions, the private first-order condition is $G_S'(\alpha) = K_S'(\alpha)$. The planner adds buyer decision benefits and relationship externalities, producing main-text equation (35). Evaluated at the private optimum, the derivative of the social objective is therefore $G_B'(\alpha) + G_R'(\alpha)$. Strict concavity places the social optimum above the private optimum when this sum is positive and below it when the sum is negative; equality leaves the two first-order conditions aligned. Boundary cases follow from the corresponding Kuhn–Tucker inequalities. QED.

## Appendix I. Empirical Identification and Preregistration-Ready Design

This appendix shows how the mechanisms can be distinguished empirically. It is a design, not a report of collected data. A confirmatory study should be registered before outcomes are generated, and power should be based on an independently labeled pilot rather than an effect size selected after inspection.

### I.1 Controlled bilateral-agent market

Human principals receive endowments and configure executable seller or buyer policies. The software, rather than an unconstrained language model, implements the registered mappings from signals and mandates to admissible actions. Markets repeat long enough for relationship state to affect later transactions. The core randomizations are: seller-signal precision, buyer-signal precision, signal evidence versus execution evidence, persistent versus resettable identity, and short versus long reward horizon. Evidence treatments must preserve information content when testing execution evidence and preserve execution guarantees when testing signal evidence.

The confirmatory hypotheses are:

1. Equal nominal surplus mandates generate less trade as absolute posterior disagreement increases.
2. Signal evidence reduces inspection in pooled-price states when the signal changes acceptance, but not under fully separating prices.
3. Execution evidence reduces rejection caused by compliance uncertainty but does not eliminate rejection caused by valuation disagreement.

4. Persistent identity and a long horizon interact positively in their effect on broad market coverage.
5. Evidence softens the buyer mandate only when it proves the proposition responsible for protective behavior.
6. Seller evidence adoption is below the surplus-maximizing level when buyer inspection and cross-user relationship benefits are not transferred.

Trade and total surplus are co-primary outcomes. Secondary outcomes are seller profit, buyer surplus, inspection, mandate strictness, broad coverage, escalation, and state-dependent switching. The analysis uses assignment indicators and their preregistered interactions, market-clustered standard errors, randomization-inference p-values for the co-primary family, and false-discovery-rate control within secondary mechanism families. Exclusions are limited to preregistered technical failures and duplicate or ineligible participation. The registration freezes the task generator, agent code hash, outcome construction, minimum detectable effect, sample size, stopping rule, and treatment labels. No language-model benchmark substitutes for incentivized principal choices.

### I.2 Structural and field identification

A field panel requires records from both agents or a protocol intermediary. The minimum schema contains committed posterior intervals or score bins, both mandate versions, price, requested and supplied evidence components, acceptance or escalation, realized outcome, counterparty identity, and subsequent counterparty choice. Exact scores need not be disclosed publicly; committed intervals and a secure research environment can preserve the inequalities used by the model.

Conditional randomization of offer overage identifies $\gamma$ from subsequent conversion or switching, while the decay of the treatment effect identifies $\rho$. Randomized identity attribution identifies $y$. A contrast between ordinary disclosure and execution attestation identifies compliance evidence, provided both arms convey the same rule description. Randomized signal evidence identifies the value of posterior alignment. The seller's horizon can be varied through continuation probability or incentive weights to identify the continuation channel in (33). A correlation between high prices and attrition is insufficient because price responds to latent value and relationship state.

**I.3 Falsification and boundary tests**

Three failures would reject central mechanisms. First, if equal mandates do not respond to posterior disagreement after holding expected value fixed, Proposition 1 lacks behavioral relevance in the studied domain. Second, if evidence that proves an irrelevant proposition changes acceptance as much as relevant evidence, the evidence-boundary mechanism is incomplete. Third, if an attributable over-cap offer has no later effect after current terms and selection are controlled, the relationship-state channel is unsupported. Conversely, evidence effects on mandate choice without effects on mechanically blocked actions would support the distinction between institutional discretion and execution restriction.